\documentclass[10pt,aps,prl,amsfonts,amsmath,amssymb,raggedbottom,longbibliography,reprint,superscriptaddress,citeautoscript,nofootinbib]{revtex4-2}
\renewcommand{\selectlanguage}[1]{}
\usepackage{graphicx}
\usepackage{dcolumn}
\usepackage{bm}
\usepackage{amsmath}
\usepackage{gensymb}

\begin{document}

\preprint{APS/123-QED}

\title{Emergent Antiphase Stacking in a Transient Charge Density Wave}

\author{Jade Stanton}\thanks{These authors contributed equally to this work.}
\affiliation{Department of Applied Physics, Stanford University, Stanford, California 94305, USA}
\author{Gal Orenstein}\thanks{These authors contributed equally to this work.}
\affiliation{Stanford PULSE Institute, SLAC National Accelerator Laboratory, Menlo Park, California 94025,USA }%
\affiliation{Stanford Institute for Materials and Energy Sciences, SLAC National Accelerator Laboratory, Menlo Park, California 94025,USA }%
\author{Ryan A. Duncan}%
\affiliation{Stanford PULSE Institute, SLAC National Accelerator Laboratory, Menlo Park, California 94025,USA }%
\affiliation{Stanford Institute for Materials and Energy Sciences, SLAC National Accelerator Laboratory, Menlo Park, California 94025,USA }%
\author{Gilberto A. de la Pe\~{n}a Mu\~{n}oz}%
\affiliation{Stanford PULSE Institute, SLAC National Accelerator Laboratory, Menlo Park, California 94025,USA }%
\affiliation{Stanford Institute for Materials and Energy Sciences, SLAC National Accelerator Laboratory, Menlo Park, California 94025,USA }%
\author{Yijing Huang}%
\affiliation{Department of Applied Physics, Stanford University, Stanford, California 94305, USA}
\affiliation{Stanford PULSE Institute, SLAC National Accelerator Laboratory, Menlo Park, California 94025,USA }%
\affiliation{Stanford Institute for Materials and Energy Sciences, SLAC National Accelerator Laboratory, Menlo Park, California 94025,USA }%
\author{Viktor Krapivin}%
\affiliation{Department of Applied Physics, Stanford University, Stanford, California 94305, USA}
\affiliation{Stanford PULSE Institute, SLAC National Accelerator Laboratory, Menlo Park, California 94025,USA }%
\affiliation{Stanford Institute for Materials and Energy Sciences, SLAC National Accelerator Laboratory, Menlo Park, California 94025,USA }%
\author{Quynh Le Nguyen}
\affiliation{Linac Coherent Light Source, SLAC National Accelerator Laboratory, Menlo Park, California 94025, USA}
\author{Samuel Teitelbaum}
\affiliation{Department of Physics, Arizona State University, Tempe, Arizona 85281, USA}
\author{Anisha G. Singh}
\affiliation{Department of Applied Physics, Stanford University, Stanford, California 94305, USA}
\author{Roman Mankowsky}
\affiliation{Paul Scherrer Institut, Villigen, Switzerland}
\author{Henrik Lemke}
\affiliation{Paul Scherrer Institut, Villigen, Switzerland}
\author{Mathias Sander}
\affiliation{Paul Scherrer Institut, Villigen, Switzerland}
\author{Yunpei Deng}
\affiliation{Paul Scherrer Institut, Villigen, Switzerland}
\author{Christopher Arrell}
\affiliation{Paul Scherrer Institut, Villigen, Switzerland}
\author{Ian R. Fisher}
\affiliation{Department of Applied Physics, Stanford University, Stanford, California 94305, USA}
\author{David A. Reis}%
\affiliation{Stanford PULSE Institute, SLAC National Accelerator Laboratory, Menlo Park, California 94025,USA }%
\affiliation{Stanford Institute for Materials and Energy Sciences, SLAC National Accelerator Laboratory, Menlo Park, California 94025,USA }%
\author{Mariano Trigo}%
\email{Contact author: mtrigo@slac.stanford.edu}
\affiliation{Stanford PULSE Institute, SLAC National Accelerator Laboratory, Menlo Park, California 94025,USA }%
\affiliation{Stanford Institute for Materials and Energy Sciences, SLAC National Accelerator Laboratory, Menlo Park, California 94025,USA }%

\date{\today}

\begin{abstract}

Photoexcitation can induce novel states in materials that are inaccessible in equilibrium, a recent example being the light-induced charge density wave (CDW) observed in LaTe\textsubscript{3}. Here, we investigate this transient CDW using infrared-pump x-ray-probe scattering at a free-electron laser, with high momentum and time resolution. We find that the transient CDW Bragg peak is broad in reciprocal space, indicating a highly disordered state. The ordering wavevector of the transient state is different from the equilibrium orders that develop in this class of materials—the transient peak appears near $(2/7, 0, 0)$ reciprocal lattice units, whereas the equilibrium $a$ order and $c$ order occur at $\approx (5/7, 0, 0)$ and $(0, 0, 2/7)$, respectively. The transient CDW is therefore distinct from the equilibrium $a$ order, differing in the relative phase of the CDW displacement between the two equivalent nearly-square Te–Te nets in the conventional unit cell. Our work highlights how photoexcitation can access states with no equilibrium analog, and how x-ray scattering can provide microscopic insight into such elusive phases.

\end{abstract}

\maketitle

Strongly coupled charge, lattice, and spin degrees of freedom in complex materials lead to rich phase diagrams~\cite{dagotto_complexity_2005} that can host hidden phases with exotic macroscopic properties that are inaccessible in equilibrium~\cite{sun_transient_2020, de_la_torre_colloquium_2021, basov_towards_2017}. Ultrafast excitation provides a route to access these elusive states; a prominent example is the recent observation of a light-induced charge density wave (CDW) in the layered materials LaTe\textsubscript{3} and CeTe\textsubscript{3}~\cite{kogar_light-induced_2020, zhou_nonequilibrium_2021}. 
Here, we use ultrafast near-infrared excitation and x-ray scattering at a free electron laser to investigate this transient state with high momentum and time resolution. Our results show that the photoinduced CDW is initially uncorrelated between layers, and evolves into a highly disordered state. From the location of the transient peak in reciprocal space, we conclude that this nonequilibrium state is distinct from the equilibrium orders in other compounds of the same class~\cite{ru_effect_2008}, differing in the interlayer stacking of the incommensurate CDW phase~\cite{gruner_density_2018}. 

\begin{figure}[h]
    \centering
    \includegraphics[width=\linewidth]{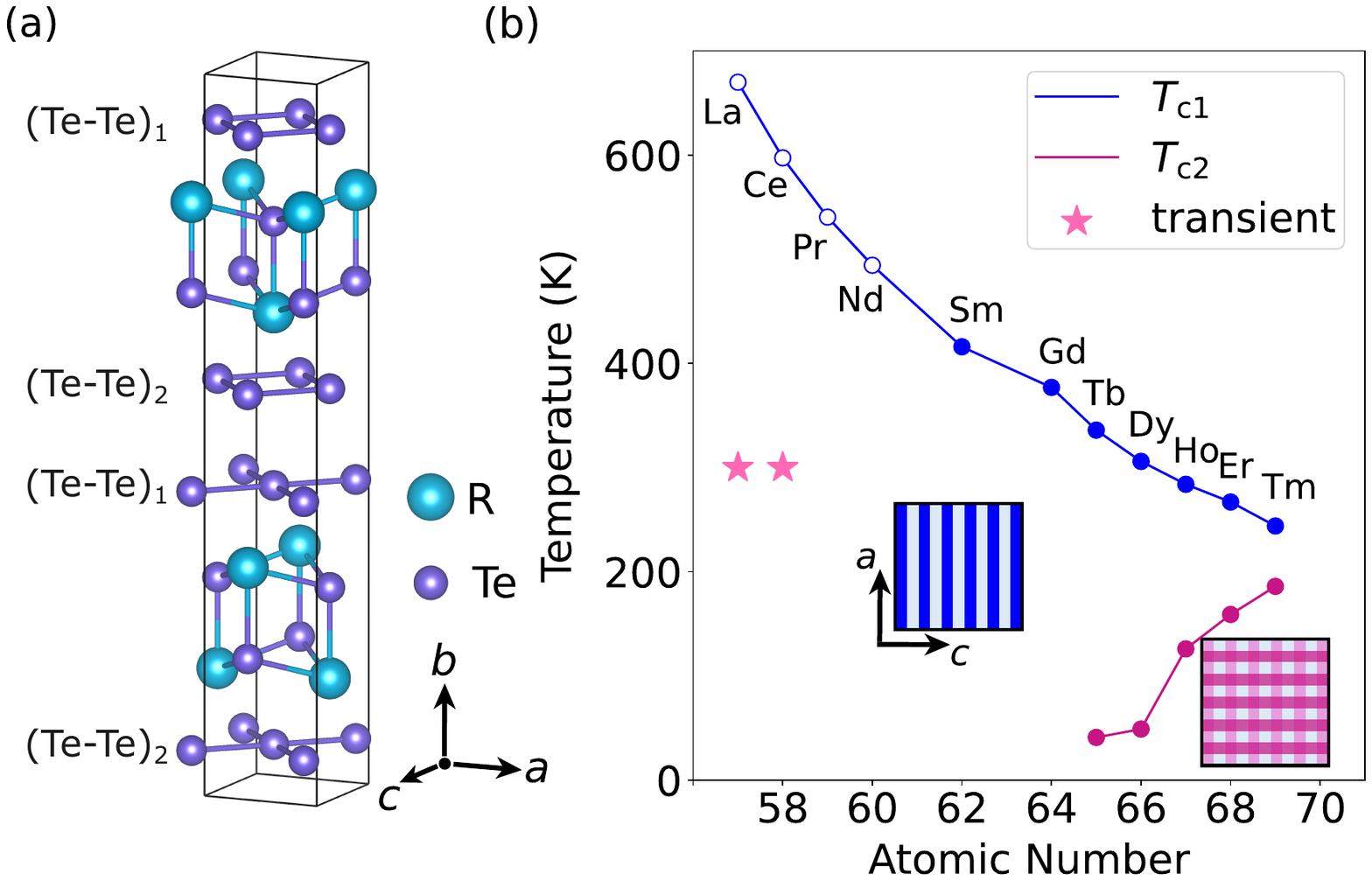}
    \caption{(a) RTe\textsubscript{3} crystal structure. Labels (Te-Te)\textsubscript{1} and (Te-Te)\textsubscript{2} denote inequivalent Te planes in the conventional unit cell. (b) Summary of the two CDW transition temperatures across the rare-earth series. Blue (purple) symbols show the $c$ order ($a$ order) transition temperature~\cite{ru_effect_2008,hu_coexistence_2014}. Open circles are extrapolated~\cite{hu_coexistence_2014,banerjee_charge_2013}. Pink stars represent the light-induced transient $a$ CDW observed here and in Refs.~\cite{kogar_light-induced_2020,zhou_nonequilibrium_2021}.}
    \label{fig:fig1}
\end{figure}

The rare-earth tritellurides ($R$Te\textsubscript{3}, where $R$ is a rare-earth ion) have attracted considerable attention due to their rich phenomena, including superconductivity~\cite{hamlin_pressure-induced_2009} and multiple, possibly competing charge orders~\cite{ru_effect_2008,banerjee_charge_2013, brouet_angle-resolved_2008, yao_theory_2006,hu_coexistence_2014,wang_axial_2022,singh_ferroaxial_2025}. Figure~\ref{fig:fig1}(a) shows the crystal structure of RTe\textsubscript{3}, which crystallizes in a base-centered orthorhombic lattice (space group \textit{Cmcm}), where the $b$-axis is the stacking direction. The CDW state is hosted by the nearly-square Te nets perpendicular to the $b$ axis, where a  weak in-plane anisotropy favors ordering along the $c$-axis. The resulting phase diagram across the rare-earth series is shown in Fig.~\ref{fig:fig1}(b). Below $T_{\mathrm{c1}}$, the system develops a unidirectional, incommensurate CDW with wavevector $\mathbf{q}_\mathrm{c} \approx (0,0,2/7)$ (reciprocal lattice units, r.l.u.)~\cite{ru_effect_2008}. For heavier rare-earth elements, a second transition occurs below $T_\mathrm{c2}$~\cite{ru_effect_2008}, giving rise to an additional CDW with wavevector $\mathbf{q}_\mathrm{a}\approx(5/7,0,0)$ r.l.u. These $c$ and $a$ orders (represented by the blue striped and purple checkerboard boxes in Fig.~\ref{fig:fig1}(b), respectively) have been shown to be highly tunable via intercalation, pressure, and strain~\cite{fang_disorder-induced_2019,straquadine_evidence_2022,singh_emergent_2024,kim_emergent_2024}. Notably, LaTe\textsubscript{3} does not exhibit the $a$-axis CDW at any temperature in equilibrium. However, recent works have shown that photoexcitation can transiently induce a CDW state along the $a$-axis in both LaTe\textsubscript{3}~\cite{kogar_light-induced_2020} and CeTe\textsubscript{3}~\cite{zhou_nonequilibrium_2021} (pink stars in Fig.~\ref{fig:fig1}(b)), which was attributed to order parameter fluctuations~\cite{zong_role_2021}. Thus far, the relationship between this photoinduced state and the equilibrium CDWs remains unclear, necessitating a direct probe of its microscopic dynamics. 

Here, we use the high temporal and momentum resolution of an X-ray Free Electron Laser (XFEL) to probe the dynamics of this transient state. We find that the transient CDW is highly disordered along the stacking direction and thus lacks full spatial coherence. Moreover, the transient peak emerges at wavevector $\mathbf{q}_\mathrm{a,tr}\approx(2/7,0,0)$, distinct from the wavevector of the equilibrium $a$ order $\mathbf{q}_\mathrm{a} \approx (5/7, 0, 0)$ observed in the heavier compounds (Fig.~\ref{fig:fig1}(b))~\cite{ru_effect_2008}. As we detail below, this difference in wavevector implies a $180\degree$ phase offset between the nearly-square Te sheets in the transient state relative to the equilibrium $a$ order. 

This experiment was conducted at the Bernina endstation~\cite{ingold_experimental_2019} at SwissFEL~\cite{prat_compact_2020}. We excited the system with a 35 femtosecond, 800 nm laser pulse with an incident fluence of 4 mJ/cm\textsuperscript{2} (unless otherwise indicated). The dynamics of the lattice response were studied by x-ray scattering using 50 fs, 10 keV x-ray pulses delayed by time $t$ relative to the excitation pulse. The sample was held at room temperature, well below the $c$ order transition temperature $T_{\mathrm{c1}}\simeq 670$ K for LaTe\textsubscript{3}~\cite{hu_coexistence_2014}.
LaTe\textsubscript{3} crystals were grown using a self-flux method~\cite{ru_thermodynamic_2006,ru_effect_2008}. The in-plane $a$- and $c$-axes of the crystal were determined by comparing the x-ray diffraction intensity of the $(0,6,1)$ peak with the forbidden $(1,6,0)$ peak. The crystal was cleaved before the experiment and kept under nitrogen gas flow to prevent oxidation. The sample normal was along the $b$-axis.
To facilitate penetration depth matching between the optical pump and x-ray probe, the experiment was conducted in a grazing incidence geometry. The x-ray beam was incident at $1\degree$ relative to the sample surface and focused to a 0.01 mm $\times$ 0.2 mm spot. The laser was incident at $5\degree$ and focused to a spot size of 0.3 mm $\times$ 0.4 mm. Scattered photons were collected by a Jungfrau area detector positioned 482 mm from the sample.

\begin{figure}[htb]
    \centering
    \includegraphics[width=0.5\textwidth]{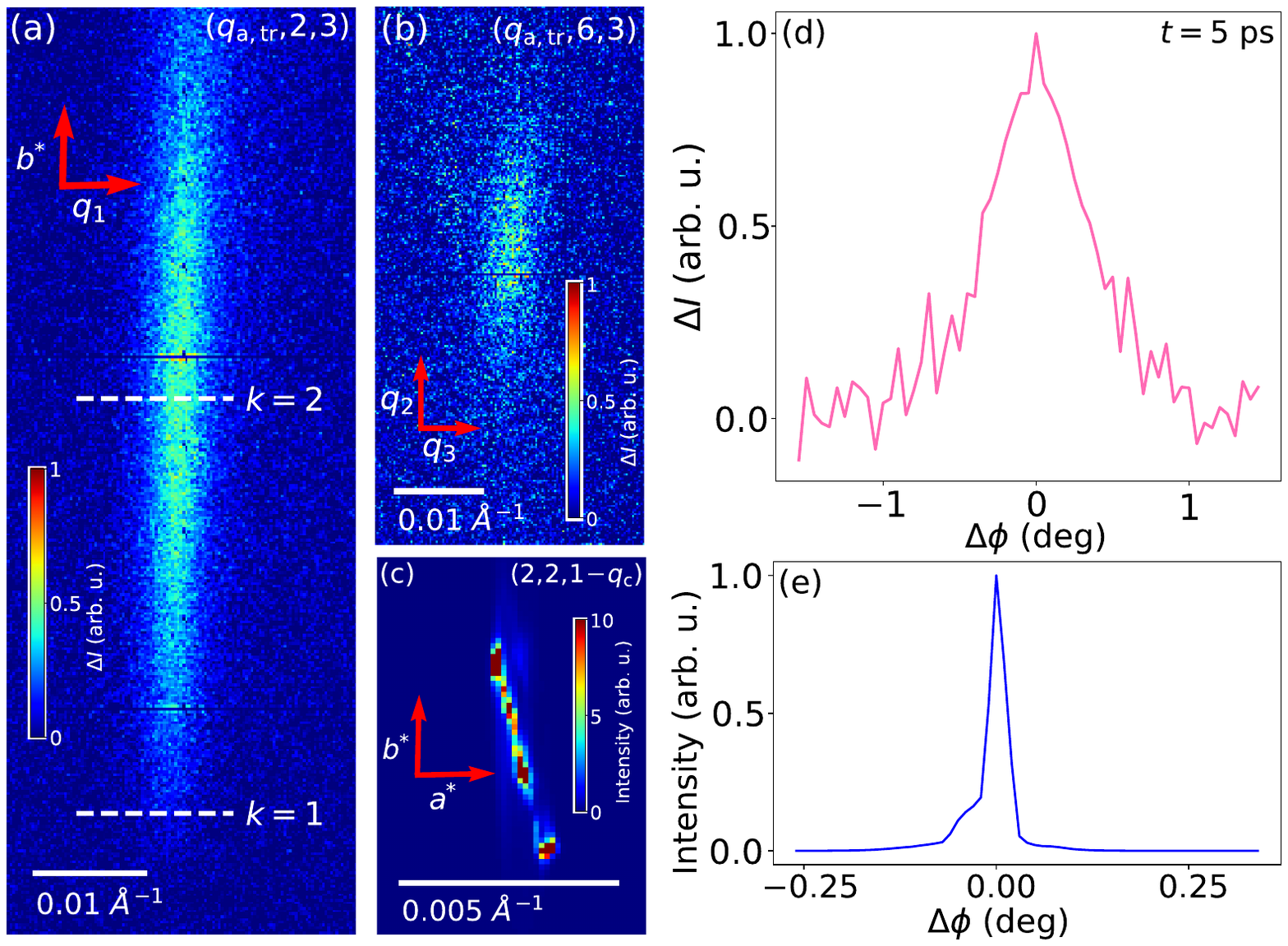}
    \caption{Intensity change $\Delta I(\mathbf{q},t)$ integrated over pump probe delays up to 4 ps around (a) $(q_{\mathrm{a,tr}},2,3)$ and (b) $(q_\mathrm{a,tr},6,3)$ in reciprocal space. The red arrows indicate directions in reciprocal space, where $\mathbf{q}_1\approx 0.50a^* + 0.87c^*$, $\mathbf{q}_2\approx-0.03a^* + 0.99b^* - 0.16c^*$, and $\mathbf{q}_3\approx 0.47a^*+0.15b^*+0.87c^*$. (c) Intensity $I_{\mathrm{off}}(\mathbf{q},t)$ around the $(2,2,1-q_{\mathrm{c}})$ equilibrium $c$ order peak. (d,e) Intensity integrated over the (d) $(q_\mathrm{a,tr},2,3)$ transient and (e) $(2,2,1-q_\mathrm{c})$ equilibrium CDW peaks as the azimuthal angle $\phi$ is rotated along the sample normal. To capture the maximum intensity of the transient peak, the incident pump fluence was 8 mJ/cm\textsuperscript{2} and the time delay between the pump and the probe pulses was $t=5$ ps. 
    } 
    \label{fig:fig2}
\end{figure}

\begin{figure}[htb]
    \centering
    \includegraphics[width=\linewidth]{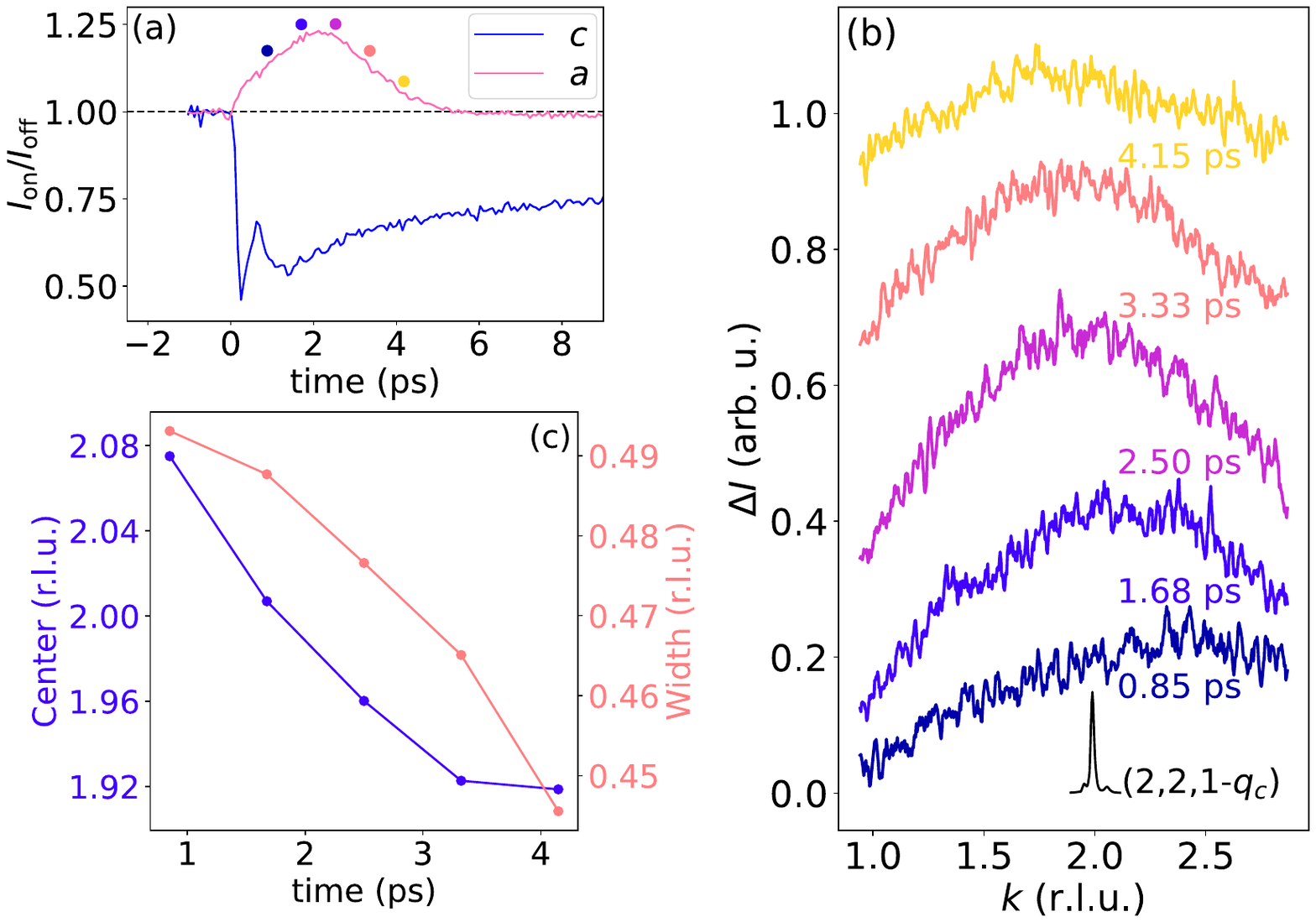}
    \caption{ (a) Relative integrated x-ray intensity as a function of pump-probe delay for the $c$ (blue) and $a$ (pink) CDWs shown in Figs.~\ref{fig:fig2}(a) and (c), respectively. Both traces are normalized by the equilibrium intensity. (b) Intensity change $\Delta I(\mathbf{q})$ of the $(q_{\mathrm{a,tr}},3,2)$ peak along the $b$ direction for representative time points indicated by colored circles in (a). Traces are vertically offset for clarity. The black trace shows the intensity of the $(2,2,1-q_\mathrm{c})$ peak. (c) Center of mass (blue) and standard deviation (pink) of the intensity profiles in (b) as a function of time.}
    \label{fig:fig3}
\end{figure}

Figures~\ref{fig:fig2}(a) and (b) show the detector image of the x-ray intensity change $\Delta I(\mathbf{q},t) = I_\mathrm{on} - I_\mathrm{off}$ near (a) $(q_\mathrm{a,tr},2,3)$ and (b) $(q_\mathrm{a,tr},6,3)$. Here, $\mathbf{q}$ is the momentum transfer and $I_\mathrm{on}$ and $I_\mathrm{off}$ denote the x-ray intensity with and without photoexcitation, respectively. The signal was integrated in the range $0 < t < 4$~ps. These two slices of momentum provide different cross sections of the transient peak and demonstrate its finite extent in three dimensions. The increased x-ray scattering intensity after excitation indicates the development of lattice distortions with these wavevectors. The extent of this feature in reciprocal space provides information about the ordering of the CDW distortion within the crystal: the broad extent of the peak in the $b$ direction (indicated by the red arrow labeled $b^*$ in Fig.~\ref{fig:fig2}(a)) indicates a short correlation length in the out-of-plane direction. For comparison, Fig.~\ref{fig:fig2}(c) shows the x-ray intensity before laser excitation $I_\mathrm{off}(\mathbf{q},t)$ around the equilibrium $(2,2,1-q_\mathrm{c})$ superlattice peak. Compared to both Figs.~\ref{fig:fig2}(a) and (b), the equilibrium $c$ order peak in (c) is well-localized in reciprocal space (see the difference in scale bars). 
Despite its broad extent, the transient peak is a true diffraction peak from a weakly ordered lattice, as confirmed by its sensitivity to rotation around the sample azimuthal angle as shown by the rocking curve in Fig.~\ref{fig:fig2}(d). The corresponding rocking curve for the equilibrium $c$ order peak is shown in Fig.~\ref{fig:fig2}(e).  
 
Figure~\ref{fig:fig3}(a) shows  $I_\mathrm{on}/I_\mathrm{off}$ integrated over the diffraction peak as a function of delay for the equilibrium $c$ (blue) and transient $a$ (pink) CDWs. Following photoexcitation, the $c$ order intensity decreases as the pre-existing CDW distortion is suppressed~\cite{moore_ultrafast_2016,trigo_coherent_2019,zong_evidence_2019,trigo_ultrafast_2021,orenstein_dynamical_2025}. In contrast, the transient CDW exhibits an initial rise and reaches a maximum at 2 ps before being fully suppressed by 5 ps~\cite{kogar_light-induced_2020,zhou_nonequilibrium_2021}. 

 To further elucidate the dynamics, we examine the out-of-plane correlation length during the formation and suppression of the transient peak. Figure~\ref{fig:fig3}(b) shows a cut through the center of the peak shown in Fig.~\ref{fig:fig2}(a) along the $b^*$ direction at several time delays, as indicated by the colored circles in Fig.~\ref{fig:fig3}(a).  The different time delays are displaced vertically for clarity. For comparison, the black trace shows an equivalent cut through the equilibrium $c$ order peak shown in Fig.~\ref{fig:fig2}(c). The transient peak is significantly broader than the $c$ order peak and exhibits a shift in its center and sharpening over time. To quantify these changes, we plot the center of mass and standard deviation of each intensity profile, shown in Fig.~\ref{fig:fig3}(c) in blue and pink, respectively. As the transient intensity reaches its maximum, the peak center approaches $k=2$ r.l.u. The peak also continuously narrows along $k$, indicating an increase in the out-of-plane correlation length.

\begin{figure}[h]
    \centering    
    \includegraphics[width=\linewidth]{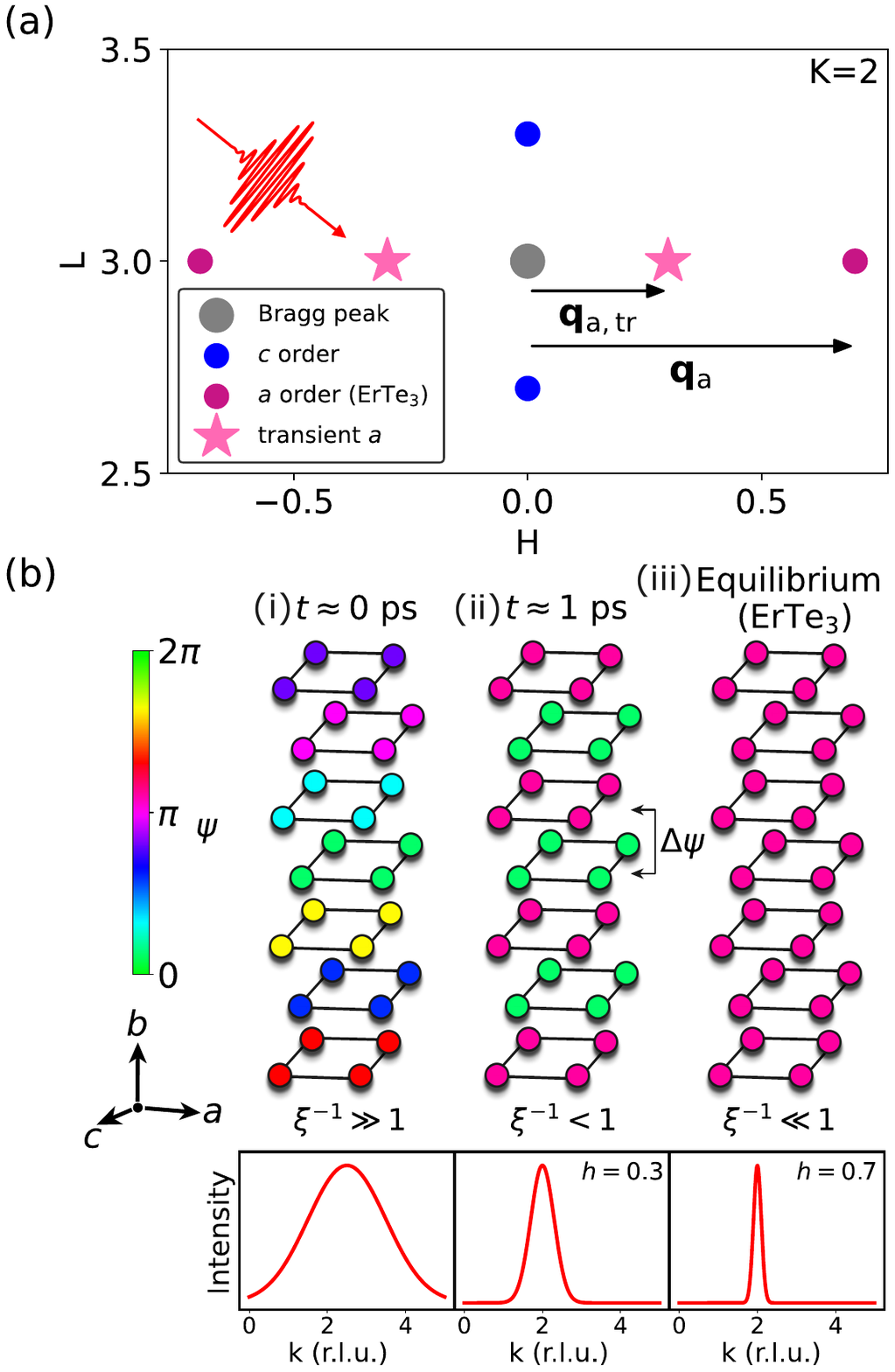}
    \caption{(a) Schematic showing the location of the different features in reciprocal space centered around $(0,2,3)$. (b) Illustration of the relative phases of the (Te-Te)\textsubscript{1} planes over time in the transient CDW state, compared to the equilibrium $a$ order. (i) At early times, the correlation length $\xi$ is short in the stacking axis, and the phases of the (Te-Te)$_1$ planes across primitive unit cells are random. (ii) At time $t\approx 1$ ps, the correlation length has increased, so that neighboring (Te-Te)\textsubscript{3} planes exhibit a $180\degree$ phase offset relative to the equilibrium case (iii). Bottom plots show a schematic of the resulting behavior of a CDW peak along the $k$ direction in each of these cases.} 
    \label{fig:fig4}
\end{figure}

Both the peak position and width provide key insights in the nature of the transient state. Figure~\ref{fig:fig4}(a) summarizes the different CDW states in RTe\textsubscript{3} in reciprocal space. It shows the $k=2$ plane centered on the $(0,2,3)$ reciprocal lattice point. The photoinduced $a$-axis CDW appears at $\mathbf{q}_{\mathrm{a,tr}}\approx(2/7,0,0)$ (pink stars), whereas the equilibrium $a$ order observed in other RTe\textsubscript{3} compounds occurs at $\mathbf{q}_\mathrm{a}\approx(5/7,0,0)$ (purple circles). Owing to the \textit{Cmcm} symmetry of the crystal, the zone boundaries of the conventional unit cell lie at $(1,0,0)$ and $(0,0,1/2)$ along $a^*$ and $c^*$, respectively. Consequently, backfolding renders $(0,0,2/7)$ and $(0,0,5/7)$ equivalent along the $c$-axis, while for wavevectors along the $a$-axis $\mathbf{q}_\mathrm{a,tr}$ and $\mathbf{q}_\mathrm{a}$ remain distinct. Consistent with this picture, prior density functional theory studies of DyTe\textsubscript{3} identified soft modes at both wavevectors along the $a$-axis, with only the $\mathbf{q}_\mathrm{a}$ mode becoming unstable below $T<T_\mathrm{c2}$~\cite{maschek_competing_2018}. Additionally, the strain induced CDW observed above $T_\mathrm{c2}$ in ErTe\textsubscript{3} also has wavevector $\mathbf{q}_\mathrm{a}$~\cite{singh_emergent_2024}. The observed transient CDW at $\mathbf{q}_\mathrm{a,tr}$ therefore represents a distinct instability from the equilibrium $a$ order CDW.

Next, we consider how these observations manifest in real space. Shortly following photoexcitation, the correlation length $\xi$ along $b$ is comparable to a single unit cell (Fig.~\ref{fig:fig3}(c)), indicating that the transient CDW is uncorrelated between the two-dimensional Te-Te planes that host it. This is described in Fig.~\ref{fig:fig4}(b)(i), which illustrates a stack of (Te-Te)$_1$ planes over several unit cells in the $b$ direction. For simplicity we show only the (Te-Te)$_1$ plane and its phase with respect to other equivalent (Te-Te)$_1$ planes; the (Te-Te)$_2$ planes exhibit similar behavior across adjacent primitive unit cells (see Fig.~\ref{fig:fig1}(a) for the definition of the planes). The color of each plane represents the phase of the transient CDW compared to the equilibrium CDW, $\psi$. Shortly after excitation, $\psi$ is uncorrelated between the layers resulting in a broad diffraction peak along $b^*$ as illustrated in the bottom panel. Around this time, modulations with wavevectors $\mathbf{q}_\mathrm{a,tr}$ and $\mathbf{q}_\mathrm{a}$ are indistinguishable because the Brillouin zone boundaries of a single two-dimensional Te-Te layer are located at 0.5 r.l.u. in both the $a^*$ and $c^*$ directions. The distinction only emerges as interlayer correlations develop at later delays and the three-dimensional crystal structure becomes relevant. The transient CDW state acquires a relative $\Delta\psi=\pi$ phase shift between adjacent (Te-Te)$_1$ planes compared to the equilibrium CDW, as illustrated in Fig.~\ref{fig:fig4}(b)(ii) and (iii). The  $\Delta\psi=\pi$ stacking results in destructive interference at $\mathbf{q}_\mathrm{a}$ and the emergence of the peak at $\mathbf{q}_\mathrm{a,tr}$.

We presented results on ultrafast x-ray scattering with high momentum and time resolution of the photoinduced $a$ CDW in LaTe\textsubscript{3}. We found that this light-induced transient state is highly disordered, particularly in the $b^*$ direction perpendicular to the nearly-square Te-Te nets. 
We further showed that this peak appears centered at a wavevector distinct from the equilibrium $a$ order found in other rare-earth tritellurides, demonstrating that the photoinduced state is a new (albeit disordered) transient phase that does not exist in equilibrium. These results demonstrate the capability of momentum-resolved x-ray probes to elucidate the microscopic structure of transient states and establish a pathway for identifying and controlling these states in driven quantum materials~\cite{basov_towards_2017}.

We acknowledge the Paul Scherrer Institute, Villigen, Switzerland for provision of free-electron laser beam time at the Bernina instrument of the SwissFEL ARAMIS branch. J. S., G. O., R. A. D, G. A. P. M., Y. H., V. K., D. A. R., and M. T. were supported by the U.S. Department of Energy, Office of Science, Office of Basic Energy Sciences through the Division of Materials Sciences and Engineering under Contract No. DE-AC02-76SF00515. Crystal growth and characterization was supported by the Department of Energy, Office of Basic Energy Sciences, under Contract No. DE-AC02-76SF00515 (A. G. S. and I. R. F. ). G. O. acknowledges support from the Koret Foundation. R. A. D. acknowledges support through the Bloch Postdoctoral Fellowship in Quantum Science and Engineering from the Stanford University Quantum Fundamentals, Architectures, and Machines initiative (Q-FARM), and the Marvin Chodorow Postdoctoral Fellowship from the Stanford University Department of Applied Physics. Q. L. N. acknowledges support by the Q-FARM Bloch Fellowship and U.S. DOE (DE-AC02- 76SF00515)

\bibliography{bib_v4}

\end{document}